# "Voice over IP in the WiFi Network business models: Will voice be a killer application for WiFi Public Networks?"

Jorge Infante, Boris Bellalta

Research Group on Networking Technology and Strategies

Universitat Pompeu Fabra

Email: {jorge.infante, boris.bellalta}@upf.edu

P. Circumval·lació 8, 08003 Barcelona (Spain)

Phone/Fax: +34935422745 / +34935422517

*Abstract*

*The stunning growth of WiFi networks, together with the spreading of mobile telephony and the increasing use of Voice over IP (VoIP) on top of Internet, pose relevant questions on the application of WiFi networks to support VoIP services that could be used as a complement and/or competition to the 2G/3G cellular networks. Voice has been, is, and will surely be one of most important killer applications for many of the telecommunications networks. Till now, WiFi networks are centred on Internet access services, not covering a significant market for voice services. However, the succeeding VoIP services working on top of Internet as the ones provided by Skype or Vonage, and the irruption on the terminal market of intelligent dual WiFi/Cellular phones can enable different new scenarios for VoIP over WiFi. In these scenarios, private residential and corporate WiFi connections, WiFi hotspots and wide-area networks may act as complementary infrastructure support for supplying itinerant voice services combined with instant-messaging and multimedia services at advantageous prices, and enabling new competing business models in the voice market. The paper explores the state of the art on the capability of WiFi networks to support voice services in a user itinerant context, identifying actual and future strengths, weakness, opportunities and threats. First of all, the paper reviews the key aspects on the changing structure of the voice market, dominated by a clear evolution to mobility, itinerancy and increased use of VoIP-based services. The evolution of business models, coverage for WiFi networks (hotspots, municipal networks, cooperative networks and) and user voice service use patterns are also analysed in order to asses the potentialities for complementing or -in some scenarios- substitute the cellular networks. In order to identify the critical success factors for the application of WiFi networks to carry voice communications, a multidisciplinary approach is taken, considering different aspects implied in the success of WiFi voice services. The main aspects analyzed consist on: technical issues on access technology (QoS concerns in existing WiFi networks and future evolution), terminal equipment technology and market, roaming and seamless integration of access technologies, and regulation of VoIP services. Next, the market structure is analysed, together with the cooperation/competition dynamics among different actors as cellular operators, wireless ISPs, VoIP service providers, or terminals and software providers. Several possible evolution scenarios are presented, ranging from cellular and integral operators controlling the vertical market, where WiFi hotspots are used as a complement to 3G/2G networks, to collaboration among different actors (WISPs, fixed operators using residential broadband connections and WiFi access points, VoIP Service providers, application developers, terminal equipment providers, municipalities and user communities deploying WiFi infrastructure).*

# 1. Introduction

In the last decade the deployment and use of WiFi networks in both the private and public use have grown at high rates. Although the original intended use of this technology was the deployment of wireless local area networks for data communications between computers, its low cost of deployment for small geographical areas and favorable regulation for unlicensed spectrum use have encouraged new actors in the telecommunications panorama to provide services open to the general public, testing new business models.

Most of the airports and hotels, many cafeterias, and other typical itinerant locations have deployed hotspots where the potential users can access Internet services via WiFi in a convenient and easy way to contract, or are even supplied free. In March, 2008, Jiwire, the most comprehensive commercial hotspots directory, accounted for around a quarter of a million (241,000) commercial hotspot locations around the world (www.jiwire.com).

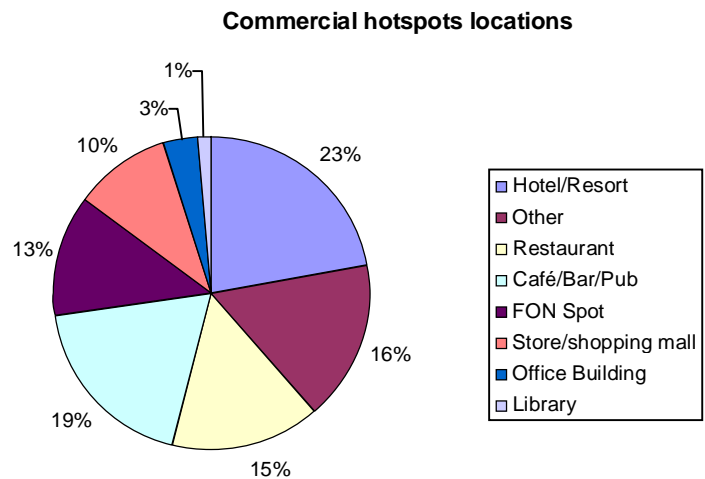

Figure 1: Commercial hotspots locations, March 2008
(source: www.jiwire.com)

Municipalities in both, North America and Europe are increasingly involved in metropolitan WiFi networks deployments oriented to supply ubiquitous services in the commercial and business areas for citizens, visitors and internal municipal use. Although the regulation issues on local administration participation has constrained municipal WiFi initiatives and still there is not a clearly defined sustainable exploitation model for these public initiatives, many municipalities are deploying citywide or hotzones under different business models, as shown in Figure 2 for the USA.

|  | JULY 2005 | FEB 2006 | APR 2006 | JUN 2006 | SEPT 2006 |
|---|---|---|---|---|---|
| Region/Citywide | 38 | 56 | 58 | 59 | 68 |
| City hotzones | 22 | 29 | 32 | 32 | 43 |
| Municipal or public safety use only | 28 | 32 | 35 | 35 | 35 |
| Planned deployments | 34 | 59 | 69 | 121 | 135 |
| TOTAL | 122 | 176 | 194 | 247 | 281 |

**Figure 2: Municipal WiFi initiatives in US**
(source: [Muniwireless, 06])

Last, but not least, residential hotspots aggregators are succeeding in involving fixed broadband users in deploying WiFi access points and sharing bandwidth between the private and public use open to the users. FON (www.fon.com/en), a spanish residential hotspots aggregator operating worldwide and funded among others by Google and Skype, claimed in its first anniversary in February 2007 to control 100.000 residential access points with 250.000 users adhered.

This growth in WiFi hotspots and metropolitan deployments has configured a panorama of widespread WiFi access availability in the most convenient locations of major cities. Although, contrary to the cellular service, the WiFi coverage is patchy and the service is provided by different operators in each location, WiFi hotspots has gained acceptance for accessing data services for business travelers.

WiFi is usually used nowadays mainly to access data services as e-mail, web browsing and other Internet-based services. In fact, the typical terminal equipments used to access WiFi services are laptops. Laptops are data services-oriented equipments where the use of voice services is not as convenient, fast and easy to use as the cellular phones. In the last year handheld terminals markets has evolved significantly, specially on the so-called "intelligent phones", provided with dual cellular (GPRS, EDGE, UMTS, CDMA) and WiFi access capabilities, allowing users to use both technologies to access the voice and data services. Although these dual terminals are still expensive and oriented to the professional and high-end market segments, presumably the prices will drop in the next years, and the 3G/WiFi terminals will spread in other lower market segments.

Internet traditional services, the ones accessed usually by the WiFi users are growing and will continue growing as more citizens will incorporate to the knowledge society: more contents and services will be available and more companies will introduce its use in their core business. However, although the increasing use of Internet services for accessing contents, these services are not the most used ones in the context of human communication. If we take a look on the most demanded and cash-flow generating telecommunication services, unsurprisingly we discover that voice has been, is, and probably will be in the next years the real killer application for both the fixed and cellular networks [Odlyzko, 04]. Figure 3 depicts data for several countries, showing how mobile and fixed services (mainly voice) revenues account for most part of the telecommunications market.

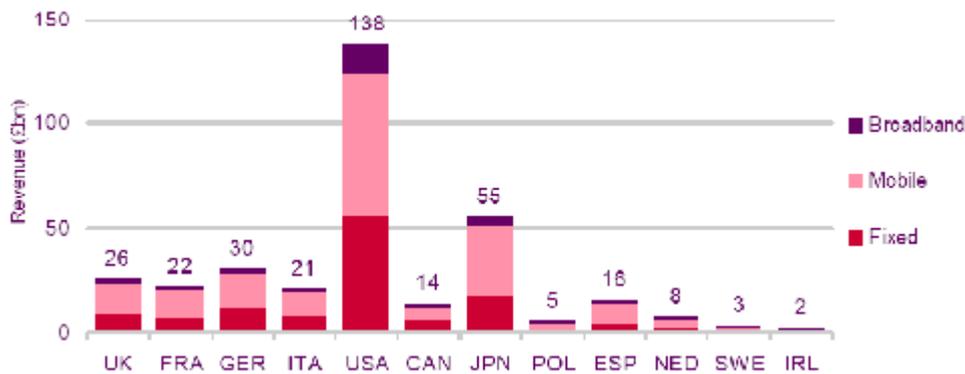

**Figure 3: Telecom Services revenues, 2006**
(source: [Ofcom, 07a])

The mobile market has experimented a very important grown in the last decade, accounting for a 13,4% share of the total telecommunication market in OECD countries in 1998, and growing up to a 39,1 % share in 2005 [OCDE, 07]. Although data and other non-voice services are increasingly used in cellular networks, voice services continue to be the largest component of mobile revenues. This is the case, for example, for the United Kingdom, one of the more developed markets for itinerant data services in cellular networks, where in 2005 voice services accounted for 79,6% of the total cellular revenues, getting a similar share (75,4%) in 2006 [Ofcom, 07a]. So, it can be said that nomadic voice is the more revenue generating service in the telecommunications market.

The aforementioned evolution of the handheld devices and intelligent phones market to incorporate WiFi as access technology, together with the WiFi standards evolution, and widespread WiFi public access, configure an opportunity window for Wireless Internet Service Providers (WISPs) to get a share on the succulent voice market. In short, if WISPs and hotspots operators are able to get even a small share of the person-to-person nomadic voice market, their revenues and strategic position could be significantly improved in the telecommunication market.

The objective of this paper is to explore the maturity grade of the WiFi technology, market structure, actors' interest, capabilities and strategies, in order to identify strengths, weaknesses, opportunities and threads for the use of WiFi public networks in voice transmission. We also analyze feasible evolution scenarios for competence and complementarity of WiFi and cellular networks and operators. The ultimate aim is to identify market opportunities, if any, for using WiFi as a support technology to transmit voice and the possible business models and companies strategies to enable this market opportunities.

## 2. What the user expects from WiFi supported voice?

Today, the major part of personal VoIP calls using Internet are motivated by its free (e.g. two users connected to Internet using Skype) or very low cost (e.g. using the SkypeOut service to call an standard phone number), specially suitable for sporadic international calls. These users, however, already accept a worst voice quality than traditional fixed telephony or cellular, caused by the bandwidth fluctuations of the IP network. For other type of calls (e.g. local calls) the benefits (relation prize/voice quality) of using VoIP probably do not compensate enough. Therefore, what requirements an user would expect to be encouraged to use the VoIP service in his every day calls, including both pleasure and business calls, for example if he owns a dual-phone with both cellular and WLAN interfaces?

There are three main expectations for a successful VoWLAN (VoIP over WLAN) service: i) *a cheaper service than the cellular one,* ii) *good voice quality (comparable to the cellular one)* and iii) *easy of* *use*. Additionally, extended multimedia services associated to the VoIP calls such as video conference, chat, file transfer, or other related multimedia contents can be also considered as means of encouraging the use of the VoWLAN capabilities.

The first expected requirement for VoIP is a cheaper voice service. Easily, users will accept to pay for a broadband Internet access (at home, at hotels, at the airport, etc.) although they normally will expect that all the IP services are free of charge. If VoIP can be seen as a special service (a dual phone with VoWLAN capabilities can help to this point) the user can accept also to pay for the calls, but at lower rates than traditional telephony.

A second requirement is a similar voice quality than the achieved by cellular networks. To evaluate the voice quality, the MOS (Mean Opinion Score) parameter can be used. It provides a scale between 1 (bad) and 5 (very good) to classify the voice quality perception. To achieve a good VoIP quality, it should be between the 4.3 value from fixed telephony and those from the cellular quality, which range around 2.9 and 4.1. The MOS value given for an specific VoIP codec decreases with packet losses and delays, which means that to provide a good VoIP service some "bandwidth" (network resources) must be guaranteed to be assigned to the call.

Finally, the third requirement is "easy of use", in terms of the complexity to make / receive calls. These procedures should be so simple as are in cellular networks (just press a few buttons) to answer to a call request, to select the destination number, etc. Ideally, the end user should ignore what technology is using to make / receive calls. This means to have a single number, to have a software capable to select the most properly network (based on the user policy) and to allow the required signalling procedures between all involved parts to simplify the different authentication and accounting issues. Another question is related to the battery duration, which for today WLAN devices is still shorter than the usual for cellular phones. Continuous use of VoWLAN can drain the battery duration of any dual phone, reducing its autonomy. However, current WLAN phones achieve already battery duration values ranging from several hours in conversation to several days in stand-by mode, which makes the use of VoWLAN feasible.

## 3. Technical aspects involved in supporting Voice over WiFi

The capacity of a hotspot (WLAN cell) in terms of number of supported calls, as well as the quality of the voice transmission over the wireless link under different channel conditions, are crucial for deciding whether this technology can be widely deployed and accepted for voice. In spite of all the research efforts in this area, there are still unsolved issues concerning the quality of VoIP calls, most commonly caused by the specific wireless network characteristics - listed below - due the fact that WLANs were only designed for data transmissions ([Bellalta et al., 06], [Sfairopoulou et al., 08]):

- **Unfairness between uplink and downlink streams**: the VoIP capacity is limited by the bottleneck effect for downlink flows in the Access Point.
- **High protocol layer overheads and ARQ mechanisms**: the number of VoIP calls is affected by the requirements to ACK all transmitted packets.
- **Fast VoIP degradation in presence of TCP flows**: the greedy and bandwidth hungry TCP flows tend to starve VoIP flows.
- **Variable capacity due to multi-rate transmissions**: mobile stations change randomly its transmission rate, which causes sporadic low capacity peaks, affecting all active calls, which can cause to drop all calls.
- **Slow and complex hand-off between APs**: the current mechanism to associate to a target AP prevents about a seamless roaming as the active VoIP calls usually are dropped during the process.

The first four impairments can be partially solved using the IEEE 802.11e [IEEE, 05] standard. It was released at the end of 2005 to fulfill the requirements of traffic differentiation and QoS provision in WLANs and has been implemented in commercial products under the WMM (Wireless MultiMedia) denomination (http://www.wi-fi.org/). These new enhancements are able to provide traffic differentiation by classifying the packets in different categories and giving them different channel access priorities by considering different MAC (layer-2) parameters for each one. These traffic differentiation enhancements are able to reduce the negative impact of best-effort traffic (predominant Internet traffic) over the VoIP calls in the access network, without regarding about the core network (which today, it's not a problem due the overprovision of resources). Additionally, the IEEE 802.11e standard also includes other mechanisms, oriented to provide VoWLAN, such as the option to suppress the ACKs, which mitigates the problem of large protocol overheads at the cost of higher loss probability and a lower voice quality. Other solutions to enhance WLANs for VoIP include call admission control and intelligent APs, which are able to manage the end-to-end QoS for VoIP calls at the same time that optimize the WLAN performance [Sfairopoulou et al., 08].

Other important issue is the WLAN coverage. It's not a pure technical question as it may be solved by placing more APs in an specific area. However, there are several problems associated with how to interconnect the APs to Internet (using a backhole cable network or a wireless mesh network) and how to assign the frequency channels to each AP (in the case of mesh networks, typical commercial products use the 2.4 GHz freq. band to communicate the clients with the AP and the 5 GHz band for AP to AP communication). Then, probably is more an economical question than a technical one. In [Arjona, 07] the authors evaluate the Google Muni WiFi network as an alternative to cellular networks for voice. They conclude that it is technically possible to support wide integral coverage but a very dense hotspot network is needed which could be equivalent to the cost of a cellular network, which still provides a better voice quality. Another issue to solve is the fifth problem, which is related to the mechanism to make hand-offs between APs. There are multiple proposals based on the Inter Access Point Protocol (IAPP) to minimize the hand-off delay and packet losses when a handoff is done [Samprakou, 05] but, there isn't a real solution implemented in commercial APs as, up to date, it was unnecessary due the very low mobility in WLANs (which results in infrequent hand-offs) and the high tolerance to those delays/looses of traditional Internet applications such as Web browsing or e-mail transfer.

Finally, we should remember that WLANs provide a pure link layer-2 network to transport IP packets. Then, all signalling solutions to couple WLANs with mobile and fixed operators in order to provide authentication, authorization and accouting are independent of the access technology. The integration with cellular network, in the so called FMC (Fixed Mobile Convergence) has several initiatives such as UMA (Unlicensed Mobile Access, [Uma Technology, 06]), iWLAN (Internetworking WLANs) or IMS (IP Multimedia Subsystem) [Munasinghe, 08]. These solutions are also able to provide transparent roaming to the user and seamlessly transitions between several access technologies.

Then, the technical issues won't be an obstacle for the VoIP success as WLANs are evolving to fully support it in both terms of QoS and coverage. However, it should be consider that existing deployed APs probably do not incorporate the enhanced capabilities to guarantee the required QoS and hand-off capabilities. To replace existent APs, the best option – in the short/mid-term - is to consider the new IEEE 802.11n/e standard, which is based on advanced PHY-layer techniques, allowing extended coverage area, achieving transmission rates up to 600 Mbps, incorporating also all IEEE 802.11e enhancements at MAC layer [Wi Revolution, 07].

# 4. WiFi Networks market structure

The public WiFi networks market is a complex and diverse ecosystem, being probably the telecommunications market where more different, small size and local actors participate in the configuration of the services offer, combined with an important participation of the traditional telecommunication operators.

The offer for commercial hotspots is provided by several different actors: first of all, *traditional fixed and cellular operators* as T-Mobile, Telefonica, or Telia-Sonera. These operators have deployed their own hotspots networks in the countries where they operate, arranging also roaming agreements with fixed and cellular operators in other countries in order to extend the WiFi access service. As shown in next table, these operators accounts for most of the commercial hotspots in many of the most developed telecom markets.

| Country | WiFi Operator | Commercial launch | Hotspots Q1 2006 | Share of total hotspots in country |
|---|---|---|---|---|
| UK | The Cloud | May 2003 | 6,600 | 62% |
| France | Orange France | Feb 2003 | 8,150 | 83% |
| Germany | T-Mobile/T-Com | Nov 2002 | 6,550 | 76% |
| Italy | Tin.it | Mid-2003 | 771 | 44% |
| USA | Wayport | Jan 2000 | 7,682 | 29% |
| Japan | NNT West | Jun 2002 | 3,030 | 45% |
| China | China Mobile | Mar 2003 | 2,100 | 69% |
| Spain | Telefonica | Jun 2003 | 1,323 | 58% |
| Netherlands | KPN HubHop | May 2003 | 720 | 37% |
| Sweden | TeliaSonera HomeRun | Oct 1999 | 524 | 60% |
| Ireland | Eircom | Feb 2004 | 550 | 75% |

**Figure 4: Largest wireless hotspot operators by country, Q1 2006**
(source: [Ofcom, 06])

Many of these traditional operators apply unique billing schemes, comprising the charges for the different services contracted by the user, including hotspots access. By now, these traditional operators do not accomplish further services integration among different services, and the hotspot network is treated as a separate one, not supporting integrated or transparent fixed voice services on it. The billing schemes are usually time-based (typically hours or days), and no specialized voice over IP services are provided by the hotspot operator. Users can anyway access VoIP services from service providers in top of Internet, as the ones provided by Skype or Vonage, but using different identifiers (E.164 numbers or alphanumeric ids) than the ones used for the service provided on the fixed/cellular networks.

*Fixed operators* play also an important role supplying WiFi access points for private use bundled with the cable, ADSL or new generation access networks supporting broadband residential service. Then, many of the existing residential access points have been acquired applying for fixed operators offers that included the residential WiFi deployment when the broadband service is contracted. As a consequence of this bundling strategy, according to the data provided by the European Commission, at the end of 2006, 34% of the European households with Internet access at home had a WiFi access point, compared to the 27% at the end of 2005 [EC, 07]. Considering that around one third of the mobile calls are made from the user premises (home or work) ([Pujol, 05], [Merrill Lynch, 04]), these WiFi access points connected to the fixed broadband connection are an important active to enter the nomadic voice market for the fixed broadband operators.

Other important actors in the WiFi market are the *commercial hotspots aggregators*. Boingo is the one controlling more locations (around 100.000 locations worldwide). Their business model is based on arranging alliances with small local WISPs, sharing the use of the WISPs hotspots for the Boingo clients simultaneously to the local exploitation by the WISP for the local clients. Each time a Boingo client connects a local WISP hotspot, the WISP receive a part of the Boingo incomes. The billing schemes are similar to the ones applied by the traditional (fixed and cellular) operators, and no specialized Voice over IP service tailored to their clients is offered. The client can anyway access any Internet service, including the VoIP service offered by other service providers as Skype. Boingo has arranged recently an agreement with Skype for supplying restricted access Skype VoIP service in any Boingo-controlled hotspot to their clients at lower rates than the Internet access rates for these hotspots (see Skypezones service at http://skypeWiFi.boingo.com).

*Local small WISPs* deploys also a part of the existing hotspot base, although the number of controlled hotspots for each WISP is modest. Most of them arrange agreements with aggregators as Boingo or fixed/cellular operators in order to benefit the large potential user base (as is the case for example for Kubiwireless, whose hotspots network can be used by the Vodafone clients).

An special case of hotspot aggregator are the grassroots wireless communities and commercial initiatives supported on residential fixed bandwidth sharing for both, the private use and public use. *Grassroot communities* typically supply free service to the public with no QoS warranties, sometime based on mesh technologies, sometimes based on hotspots in residential premises whose location is publicized on the community website. Although most of these initiatives are weak and more oriented to experimentation than to real public service, some of them located in areas lacking of complete coberture of fixed broadband as guifi in Catalonia, Spain (www.guifi.net) or Freifunk (http://wiki.freifunk.net/Kategorie:English) in East Berlin (Germany) accounts for an important base of users and hotspots.

Residential WiFi services aggregation has also deserved the interest of commercial companies that are interested in taking profit for supplying WiFi access service based on a large base of residential hotspots provided by adhered residential broadband users. The more succeeding case of *residential hotspots aggregation* is FON, a Spain-based company controlling a similar base of hotspots as Boingo, although located in less attractive positions (manly residential locations). The agreement arranged by the users adhered to FON consists on providing access to any FON client (adhered or not), allowing FON a free access for the adhered users in the whole FON network. FON has also arranged agreements with Boingo, and is establishing other agreements with fixed operators and municipalities in order to extend their hotspots base and to consolidate as the key actor in the residential hotspots aggregation market.

Last, but not least, many *retail shops* participate in the hotspot market supplying free service at their premises in order to attract clients for other services provided at the location.

The market dynamics for hotspots is dominated by several trends:

- Fixed and cellular operators are incrementing their hotspot locations in the countries where they operate and establishing roaming agreements with other operators in foreign countries. They are and will continue to be the more important actors at national level in most countries. Their strong point is based on the convenience of combining WiFi and other service offers in just one bill for the user, and the possibilities to integrate service offers under cellular, fixed and WiFi infrastructure.

- The economies of scale present in the aggregation market configure a panorama consisting on a reduced number of aggregators acting at international scale and controlling a large number of hotspots on the most interesting locations. Boingo is the clear winner in commercial hotspots aggregations based on small local WISPs and venues arrangements, and FON in by now the more important aggregator for the residential hotspots.

- Most of the more privileged locations are occupied by existing commercial hotspots controlled by fixed or cellular operators, and large commercial aggregators. In order to extend the hotspots base, both types of actors will try to establish agreements with small retail business offering WiFi in order to incorporate these locations in their hotspots offer.

When analyzing itinerant voice services provided on top of WiFi networks, it is important to study not only the hotspots offer, but also the availability of dense and integral coverage networks that could enable a real alternative to the ubiquitous cellular 2G and 3G networks. The market offer in this case is nowadays scarce and controlled by local administrations cooperating with specialized companies as Earthlink, Munifi or The Cloud. Many *local administrations* are getting involved in these citywide or area-wide WiFi networks deployment **Error! Reference source not found.**, pursuing social benefits as increasing incorporation in the knowledge society for citizens, local development, bridging the digital divide, or improve internal and external municipal services

In a first moment, companies like Earthlink and Munifi considered these dense WiFi deployments as a profitable business, and local initiatives as the Wireless Philadelphia [Wireless Philadelphia, 05] were financed by these companies. The competition in residential broadband from fixed networks, based on price reduction and bandwidth increment offer, has reduced the clients share for these municipal networks compared with their expectations. Also, nomadic access to Internet services has not still grow enough to maintain sustainable business plans for dense WiFi deployments, as the nomadic use is still centered on the most suitable locations, the ones covered by commercial hotspots.

This situation has "cold" the impetus of these companies to finance municipal deployments and the business models are changing towards:

- Anchoring a more significant part of the incomes of municipal WiFi specialized operators on medium-term contracts for the municipal internal services.

- Increase use of the municipal network for internal services (nomadic access for officers to corporate applications, surveillance and security cameras, etc.).

- Location-based publicity models enabling "free" WiFi service for users based on exposition to publicity. This is for example the model applied by Munifi in the succeeding network in Portland.

This combination of internal use with publicity-based free itinerant access, together with the increasing use of nomadic users and the possibility to support VoIP services can enable in the medium term sustainable municipal WiFi networks.

If the municipal networks prove to be economically sustainable under these conditions, these dense and city-ubiquitous networks can play a key role to compete and complement 3G networks in the nomadic voice market.

So, as we have seen, the public WiFi networks market is dominated by a few number of fixed and cellular operators deploying hotspots in privileged locations, together with commercial and residential aggregators, that control the rest of hotspots. Succeeding grassroots wireless communities are working under conditions of fixed broadband lacks, and municipal networks have not yet found an economical sustainable business model, but in the medium term, it is possible that they will success deploying profitable wide area WiFi networks.

Still there are no dense ubiquitous public WiFi networks and, by now, the only future way to enable these deployments is a success on the municipal WiFi business models, not yet reached, except for some special cases.

## 5. Market structure in WiFi terminal equipments

Handheld devices availability is a key factor to enable WiFi VoIP services to be used in a convenient way by the final user. Although the laptop market has evolved to incorporate WiFi as a "must" and there is a trend to make lighter these equipments, a laptop is not a substitute, in a nomadic scenario, for the easy to use, light and convenient cellular phone or PDAs (Personal Device Assistants).

Although PDAs have evolved to incorporate WiFi as a common feature, PDAs lacking of cellular access capabilities, even being able to supply VoIP services, cannot act as a effective substitute of cellular phones for nomadic voice, as WiFi coverages, as seen before, are in general patchy, not supplying a integral coverage.

The most adequate handheld equipments for VoIP WiFi services are the intelligent phones that combines the light weight and size of traditional telephones with the data services management capabilities provided by laptops and PDAs. The market for these intelligent dual WiFi/cellular phones has been growing at very high rates along these last two years fueled by several factors:

- New entrants offer in the mobile phones market as the Apple Iphone or the HTC touch terminals.
- Evolution of installed phone providers as Nokia to increase the offer of mobile phones combining features of PDAs, cellular phones and WiFi/3G access technologies.
- Evolution of PDAs providers to incorporate cellular 2G/3G communications capabilities to WiFi-enabled PDAs, entering in the intelligent telephones market.
- Cellular operators subsidy strategies for terminal equipment.

Most of the existing intelligent dual WiFi/Cellular phones are acquired as part-subsidized terminals supplied by cellular operators as an incentive for the clients to contract the cellular service. In this case, the terminal is configured to connect only to the cellular provider network, but the user is free to connect any WiFi network not necessarily controlled by the cellular provider (private home, work, or any hotspot or WiFi area network).

As the prices for these intelligent cellular/WiFi phones are still high, cellular providers terminal subsidy strategy is a critical success factor for spreading the intelligent phones base. The actual trends on this market confirm the proactive strategy of the cellular operators (see for example the interest shown by most of the leading cellular operators to gain exclusivity to distribute the Apple Iphone terminal). Ironically, the cellular operators are the key actors that are helping the extension of dual intelligent phones that can enable in the future the nomadic use of VoIP services based on WiFi.

## 6. Market structure in VoIP service providers

The VoIP residential market is dominated by a few leading companies offering voice calls on top of Internet with a threefold offer: free calls among users adhered to the service, low-cost calls to telephony networks (both fixed and cellular), and numbering and incoming calls from other telephony networks. Skype and Vonage are the most succeeding companies in this market.

These service providers do not operate in the broadband access markets, nor care about the access technology used to access the service. In fact, the access network technology is transparent for the service, except for the QoS capabilities supported. As the market for handheld devices is not mature enough, till now the real use of VoIP services has been centered on fixed broadband access networks.

VoIP service providers have extended a lot the user base; Skype accounted for around 276 millions of accounts worldwide at the end of 2007 and Vonage for around 2,5 millions (a user can use more than one account). The business model for these VoIP service providers is based on supplying free service to direct communication among the provider clients on top of Internet, combined with reduced rates for telephony services, being this telephony services the most important source of incomes for the company. This business model is threatened by the flat-rate tariffs increasingly applied by the telephony operators for national calls (bundled usually with broadband services) that shrink the potential market for VoIP services.

The irruption of handheld devices equipped with WiFi can open a new market for these VoIP service providers in competition or complement with ubiquitous cellular networks. Prices for cellular calls are still significantly high and flat-rate tariffs are more the exception than the norm. So, in the next years there is an opportunity window for VoIP services using WiFi networks, if the rest of aspects as handhelds availability, quality of service, network coverage and numbering integration evolve to solve the VoIP on WiFi shortcomings.

Existing VoIP service providers are well situated to exploit this market opportunity supported on their large user base and experience in interconnections with both fixed and mobile telephony networks. As their main incomes source, interconnection with telephony networks from fixed locations, is threaten by the fixed operators flat-rate tariffs, they have also the motivation to search an alternative income source based on the competition for nomadic voice calls (a clear example is the Skype-Boingo alliance explained before).

## 7. Regulation issues

### *Regulation for WiFi public networks*

WiFi technology is a special case compared with other access technologies regarding regulation issues: in most of the countries, no spectrum use license is needed to deploy and operate WiFi networks. This situation, combined with the low cost of deployment of WiFi hotspots has stimulated the deployment of private and public use networks by non-traditional telecommunication actors as the final users, retail businesses, and municipalities.

Those actors deploying and operating WiFi networks open to public act in fact as ISPs and must fulfill all the legal considerations for information services, that although may be different for each country, in general include notification to national authorities to be included on a public operators registry, connections logging for law enforcement agencies, and users data privacy protection procedures.

WiFi hotspots are based on broadband fixed data services in order to connect the WiFi access points to Internet. These broadband services are provided by third party fixed operators, who usually restrict the residential service for private use, banning service reselling or even free service provision to other users supported on the contracted fixed residential broadband service.

Legal obligations for WiFi service providers and restrictions on the use of fixed broadband residential service to be used as transport network for WiFi services suppose an important limitation for initiatives based on residential broadband use for deploying hotspots, as residential users participating on citizen initiatives or aggregation business models as the FON one, infringe usually the regulation for ISPs and contracts with the fixed broadband service providers. Although till now the law enforcement actions have been few, a clear risk exists for these initiatives where the residential users act as ISPs.

Commercial hotspots networks and municipal ones are not in general constrained by these issues, as they usually act as ISP providers, addressing legal issues on info logging for law enforcement agencies, and contracting backhaul and transport services not restricted to internal use, but open to be used as network support for providing WiFi services.

Municipalities must also comply with the regulation on public participation in providing telecommunication services. Being the telecommunication market a liberalized one, in most of the occidental countries, regulation establishes restrictions in order to avoid disloyal competition between public and private initiatives. In the case of the USA, each state has their own regulation on municipal public telecommunication services, being not allowed in some of them, and restricted with different conditions in others. The regulation on the European Community varies on different countries, but the general principle is to restrict municipal initiatives to economically sustainable business models not based on public subventions in those areas where exists commercial private broadband services offer. As the business models for municipal networks are not yet validated for configuring sustainable initiatives, the regulation on public participation restrain municipal involvement and financing on wide area WiFi networks.

So, regarding the WiFi networks open to public use, although no license is needed for spectrum use, there are some regulation issues limiting hotspots and wide-are WiFi networks deployment: initiatives based on residential hotspots do not generally comply the legal requirements for ISP service providers regulation and many times the fixed broadband contract service used to connect the residential hotspot is not fulfilled as no public service used is usually allowed by the fixed broadband operator supplying residential Internet access service. Municipal networks are also limited by the competence enforcement regulation, not been allowed in some cases, and restricted in other cases to commercial business models not based on public subventions.

### *Regulation for VoIP services*

When analyzing voice over IP services, additional regulation issues must be considered, as these services are subjected to regulations concerning similar aspects than those applicable to other voice service:

- *Consideration or not as public telephony service*, implying obligations to provide free emergency calls and location info for these calls in order to optimize the emergency service. Location info is not easy to be provided on IP networks, as the user may connect from different locations each time, even from foreign countries where there is no regulation on location info availability

- *Interconnection to both, fixed telephony networks and cellular ones*. In order to be able to send and receive calls to/from other networks, the service provider should be able to establish interconnection agreements with the rest of voice services operators.

- *E.164 Number assignment*. In order to receive calls from other networks, VoIP users should be assigned with a regular telephone number where any caller from a cellular or fixed telephony network can direct their calls. This implies that the service provider should be able to obtain public number blocks from the regulation authorities.

- *Number portability*. A step further to increase the convenience of the VoIP services is to be able to port an existing number from one provider to another, avoiding the problems derived from changing the telephone number. Number portability has been an important factor to increment competition in the existing fixed and cellular voice markets.

These issues have been addressed in different ways by the national regulation in each country. Although there are countries where VoIP services are directly banned, or restricted to communication among the service provider's clients [Melody et al, 07], in countries where the telecommunication markets are more developed and enjoy a certain grade of competition, as the USA, Japan, South Korea or the European Community, VoIP services both in internal use and connected to telephony and cellular networks are allowed under different conditions [OCDE, 07].

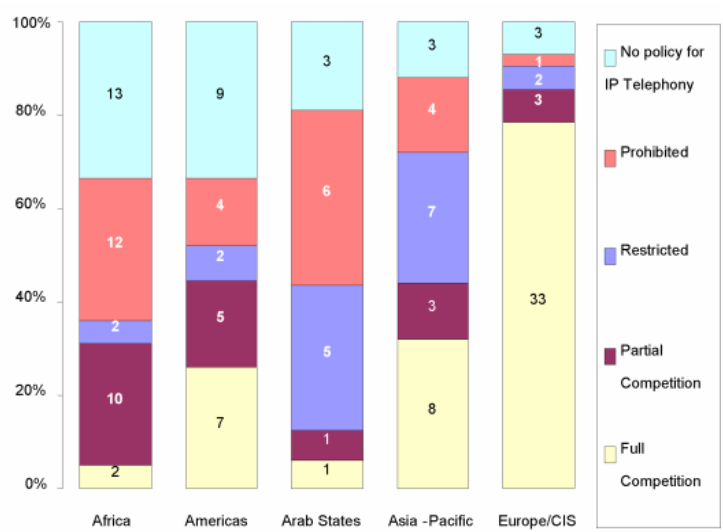

*Note: Based on responses from 149 economies. "Prohibited" means no service is possible. "Restricted" means only licensed PTOs can offer the service. "Partial competition" means non-licensed PTOs may use either IP networks or the public Internet. "Full competition" means anyone can use or offer service.*

**Figure 5: Regulation of VoIP services around the World, 2005**
(Source: ITU World Telecommunication Regulatory Database, 2005 questionnaire).

For these more mature markets, the regulation of VoIP is oriented to facilitate innovation and spreading of VoIP, based on as light as possible obligations for all the involved parts (see, for example the common position of European Regulators Group, [ERG, 07]). VoIP services are not considered in general telephony, and, as a consequence, issues as location info provision for emergency calls are not enforced or are based on a "best effort" scheme. Interconnection with PSTN and cellular voice networks is not in general enforced in regulation, but existing VoIP service providers as Skype or Vonage have managed to arrange interconnection agreements with cellular and fixed telephony operators in order to manage traffic coming from or addressed to other networks directly with other operators or through intermediate telephony operators handling the VoIP traffic exchange with other voice networks.

Numbering assignment for VoIP services is addressed in the regulation for many countries, and special numbers blocks are reserved for these services, distinguishing between two situations: geographic and nomadic numbers. The first ones are to be used in fixed locations or small geographical areas, while the nomadic ones are not constrained to be used in a reduced geographical area. Nomadic numbers resembles in some way the cellular numbering, while the non-nomadic ones are thought to be used in a VoIP fixed location service, resembling the traditional fixed telephony.

Although in some countries the regulation for VoIP services consider number portability among VoIP services (i.e., a user changing from one VoIP service provider to another can retain his number), national regulations do not address number portability between VoIP services and cellular or fixed services. As VoIP services are not being regulated as traditional voice services, and number structure is a reference for the user in order to know the rates that are to be applied to the outcoming calls, it is unlikely that in the short and medium term number portability between VoIP and other voice services will be considered in the national regulations.

So, in short, the regulation of VoIP services in the more mature markets as the USA, European community, Japan and South Korea allows the commercial use of these services. E.164 number blocks are also available to be used in the regular ways that the clients are used to for managing incoming and outcoming calls. As a limitation, the clients cannot migrate their cellular or fixed number to the VoIP service and must use a new different number for making and receiving calls.

*Network neutrality*

Network neutrality regulation is also an important issue when analyzing evolution of any VoIP service supported in third party broadband services. If broadband service providers are allowed to restrict the services provided via their access networks (no network neutrality regulation), they could forbid the use of VoIP services in order to avoid competition for the voice services. As hotspots are supported by backhaul and transport fixed networks, this ability to select the services users can access to, could disable the possibility to provide WiFi VoIP services supported on WiFi hotspots.

Network neutrality regulation will also impact on supporting VoIP services on top of cellular networks, as the cellular operators could also restrict the use of VoIP services on top of the Internet access mobile services if the regulation allows for service discrimination on access networks. As analyzed in [Felten, 06], discrimination on services applied by networks providers can specially hurt voice over IP services, both, fixed and nomadic ones. Till now, only anecdotic evidence is available on VoIP services discrimination (as the case for North Carolina ISP Madison River blocking DSL customers from using Vonage VoIP services described in [Meinrath et al., 06]).

Network neutrality regulation is now a hot topic with encountered positions between the network operators defending regulation oriented to allow services priorization and/or access control, and communication and information services providers supporting their services on Internet services provided by these traditional operators in the other side.

# 8. Market actors strategies on VoIP/WiFi

In this section we will review the interests, capabilities and limitations for participating in the VoIP WiFi market for each relevant actor on the VoIP WiFi market, in order to identify possible evolution scenarios that could enable VoIP WiFi services growth.

*Fixed operators*

Fixed operators are well situated to exploit the potential benefits of supplying VoIP on top of WiFi. They control the key resource for providing hotspots connections, the fixed access network, and many of them operate commercial hotspots. Also, fixed traditional operators lacking of cellular networks are not conditioned by the possible "cannibalization" effects of existing nomadic voice market, as they do not offer these services on their fixed network infrastructure.

The main strength for entering the WiFi VoIP market in the case of fixed operators is based on their large base of residential broadband service. They can use this infrastructure for configuring large residential hotspots service arranging agreements with the residential user to share the private use of the broadband residential service with a public use open to other users.

Although most the fixed operators have not yet applied these strategies specifically for VoIP, two collaboration initiatives between FON and fixed operators, (BTFON with British Telecom [BTFon, 07] and an agreement with Time-Warner) can be considered a first step in this direction.

Both companies have adapted their broadband residential service contracts in order to allow WiFi service supply to FON users using their infrastructure. In the case of British Telecom, the clients participating in the initiative are included in the FON community and can connect free to any other FON hotspot adhered to the community. In both cases, the goal of the fixed operator is to make more attractive the fixed broadband offer to potential users interested in getting involved in residential hotspots collaborative deployment.

Although fixed operators have not still implemented explicit WiFi VoIP strategies, their control of the fixed telephony and broadband services, combined with the advances in integration architectures as IMS and the dual WiFi/cellular phones can habilitate them enter the nomadic voice market.

The strategy in this case will probably be supported on extending the voice service supplied under a fixed number to any location where the user can connect Internet via a handheld terminal using WiFi or any other access technology supporting VoIP. Arrangements as BTFON can also enable service offers where the fixed clients can extend the use of the service outside their premises.

In any case, as pure fixed operators they do not control cellular networks and WiFi service is not ubiquitous. Under this situation, service proposals integrating VoIP over WiFi coming from fixed operators will not act as a complete substitute of cellular service, but as a fixed telephony extension to be used (receive or send calls) from places where the users can enjoy WiFi coverage, as commercial hotspots (being or not operated by the fixed operators) or the private WiFi coverage in their premises.

### *Cellular and integral operators*

***Cellular operators*** have made huge investments on 3G networks and are betting for the cellular technologies evolution as the support for future convergent networks for itinerant use (HSPA and EVDO in the medium term and LTE in the long term). Their business model, based on deployment of base stations does not consider WiFi as a key enabling technology, but as a complement to the core cellular networks in special locations.

As cellular operators today owns almost all the nomadic voice market, any collaboration with other market actors (for example, VoIP service providers) pose a potential thread for the existing market share, cannibalizing incomes in the cellular market.

This situation of large sunk investment on 3G networks combined with the control of the nomadic voice market, shared by the cellular operators, indicate that these actors will not be a first mover in the VoIP/WiFi market.

So, for the case of cellular operators there is a clear trend to cover both voice and data nomadic services using cellular technologies networks, avoiding investment on alternative WiFi access networks. The evolution of cellular networks to increase data bandwidth suppose a clear threat to existing commercial hotspots, and depending on the evolution of terminals, WiFi coverage and charges for the services in each case, high speed cellular technologies like HSPA and EVDO now, and LTE in the future can squeeze or WiFi public networks business opportunities or even corner WiFi to the initial intended use: private local area networks.

***Integral operators*** (those controlling simultaneously the fixed and cellular markets) are in a good position to combine the cellular and broadband offers using WiFi as the preferred access technology when the user is at home and cellular technologies otherwise, using unique integrated number and supporting free or reduced-rates calls at home by means of a dual WiFi/3G phone preconfigured for this use. This is the case of the integrated offer of TeliaSonera in Denmark [Observer, 06], supported on the Unlicensed Mobile Access ([Uma Technology, 06]).

Although pure cellular operators can reply integral operators strategy lowering tariffs for outgoing calls calls from the base station nearer to the user premises, integral operators can better optimize the infrastructure deployment, reducing cellular base station deployments taking profit from the fixed infrastructure use via WiFi access in the user premises (or even in other locations where the integral operator can offer WiFi connectivity based on residential bandwidth sharing or hotspots deployment).

*WiFi hotspots aggregators*

Hotspots aggregators can see their incomes incremented if they succeed on capturing part of the cellular traffic on their controlled networks. At a first glance, these actors could apply these strategies:

- Offering VoIP WiFi supported service to the hotspots clients supplying similar services to the ones provided by service providers in top of the Internet (free calls among clients, low rates for calls to other voice networks and numbering to receive calls).
- Establishing collaboration agreements with a VoIP service provider in order to get special benefits for the hotspots users (for example, lower prices for VoIP services) in order to increase the hotspots use.

As the utility of the VoIP service for the user grows as the VoIP service user base extends, the second strategy, to collaborate with large VoIP service providers is better for attracting users to the hotspot service. In fact, these are the cases with Boingo (the largest WiFi commercial hotspots aggregator) and FON (also the largest one for residential aggregation). As told before, Boingo and Skype commercialize the Skypezone service, offering low-rate access to the Skype service in Boingo locations. In the case of FON, this company has launched a WiFi phone with embedded access to Skype service through the FON network.

So, the strategy for attracting incomes from WiFi VoIP services in the case of the commercial and residential hotspots aggregators is based on collaboration with the VoIP service providers on top of the Internet offering low-rate WiFi VoIP access and promoting cheap WiFi phones configured to use the aggregator WiFi hotspots to access the VoIP service provider.

*VoIP service providers*

As seen before, VoIP service providers have difficulties to compete with the fixed telephony flat-rate tariffs for national calls and are interested in entering the nomadic voice market where the rates and consequently the margins are higher.

Their more important strengths, a large users base, availability of interconnection agreements for both fixed networks and mobile ones, and the experience in managing VoIP numbering can be used to provide a semitransparent service to the user based on providing a VoIP number as a convenient address contact when the user is under the coverage of one of its preferred WiFi networks (at home, work, commercial hotspot or municipal network). When not connected to WiFi networks, the user uses a cellular service contracted with any cellular operator.

In Figure 6 we have exemplified this kind of numbers combination and redirection using the facilities provided by Skype to manage VoIP calls. Suppose that a user has a mobile number contracted with a cellular operator (+665334321 in the example) and wants to use a dual phone and a WiFi network for outgoing and incoming calls when under coverage, and the cellular network otherwise in the more transparent and convenient way.

For this case, Skype would provide the user with a Skype VoIP number (say (203) 599-1176) that can be used for receiving incoming calls from any fixed or mobile network when the called user is connected to Internet and make calls both, to Skype users, and other networks ones. The tariffs that will be applied to both the calling and called users are computed considering that the assigned VoIP number corresponds to a local number in the city where the user has selected to assign the VoIP number. Any call incoming and originated by a Skype user is free.

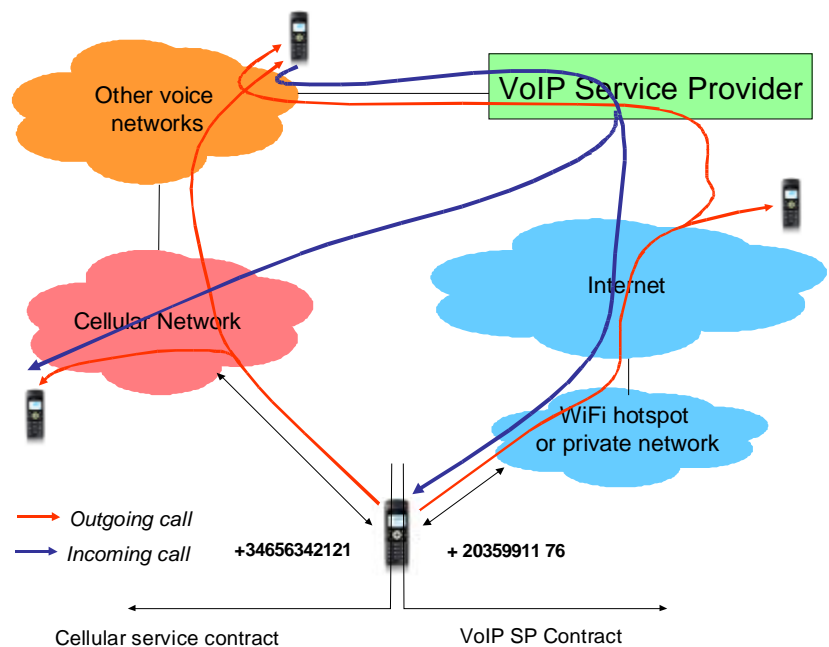

Figure 6: MVNO combined cellular/WiFi VoIP Service

When the user is not connected to Internet (the terminal is out of coverage from the preferred WiFi network), in order not to lose the call, he can use the Skype call forwarding service that will redirect the call addressed to the VoIP number to any cellular or fixed number defined by the user when he is off line the Skype service. The called user will pay for each incoming call the same fee as established for outgoing calls to cellular or fixed numbers located in the same country where the cellular or fixed number is located.

Is this a transparent service? It is not. The user have to contract services with both, the cellular and the VoIP service provider, and an additional payment is to be made to the VoIP service provider for each call forwarded to the cellular phone. Anyway, although not totally transparent, this scheme allows for:

- Free outgoing and incoming calls for both the calling and called parties for on-net calls (users of the VoIP service provider) when connected to WiFi.

- Reduced rates for outgoing calls to cellular or fixed telephony networks when under coverage of WiFi network.

- When off-line the WiFi network, the user can make calls using the mobile network, although the called party will be notified with the cellular number and not the VoIP one.

- When off-line the WiFi network, the user can receive the calls to the VoIP number using the cellular service, although he must pay an extra free to redirect the call to the cellular number.

The more the user stay on-line and the more often called and calling parties are users of the VoIP service provider, the more the user can profit the advantages of this scheme. So, provided that adequate handheld devices are available, the VoIP service will be interesting for users transiting often on locations where contracted or free WiFi is available and only occasionally receive calls in locations where no coverage is available.

New convergent architectures as IMS allows a more transparent service, supplying with just one number the service and using in a transparent way the Internet access when available and the cellular network otherwise both for outgoing and incoming calls. Of course, in this case the collaboration from both parties, the cellular and the VoIP service providers is needed to implement this kind of transport architectures. If the cellular operator is not willing to collaborate, the VoIP cannot go further in the integration model. As we have seen before, the cellular operators are not keen on collaborating with other actors to cannibalize their nomadic voice incomes. This means that VoIP service providers will depend on WiFi coverage availability provided by other actors as WISPs, fixed operators, municipalities and aggregators to put in value their nomadic services and the transparency of the service will be somehow limited.

In order to increment the service demand, VoIP service providers can establish alliances with other complementary actors as:

- Hotspots operators in order to arrange special offers to provide VoIP services as the ones analysed in the hotspots aggregators strategies (Skypezone service provided by Boingo and Skype).

- Terminal equipment suppliers in order to facilitate the configuration of handheld devices to use the provider service or even direct commercialization of WiFi Phones. Skype for example has arrangements with several suppliers supporting Skype service on their dual or WiFi phones (see for example, Netgear Skype phone in http://tools.netgear.com/skype/, or the agreement with Motorola for its SkypeIn service to be available on manufacturer handset for either WiFi or 3G networks [Pujol, 05])

As VoIP service providers are seen as a threat by the traditional cellular operators, no alliances or commercialization joint strategies are likely to be established between VoIP service providers and these traditional operators.

## *MVNOs*

Mobile Virtual Network Operators (MVNOs) are cellular operators that, instead of deploying cellular networks, use the networks deployed by other operators in order to supply mobile services profiting their own commercialization networks and contracting wholesale cellular network use. By now, these MVNOs are in general oriented to low-income market segments, offering lower rates than traditional operators and supporting their service offering on bundling and supporting integrated billing with other services or products. Most MVNOs support their service completely on the signaling network and service management facilities provided by a cellular network operator and their contribution to the value chain is centered on using their commercialization network for client management [Dippon et al., 06].

In the context of WiFi supported VoIP services, a MVNO deploying signaling and service management facilities could support its nomadic voice service in both, a cellular network and Internet, supplying nomadic voice service via VoIP, on top of WiFi when the user is connected to Internet (by means of a public WiFi hotspot or a municipal network –case A in the figure- or user's private WiFi service, case B in the figure) and the third party cellular network otherwise (case C).

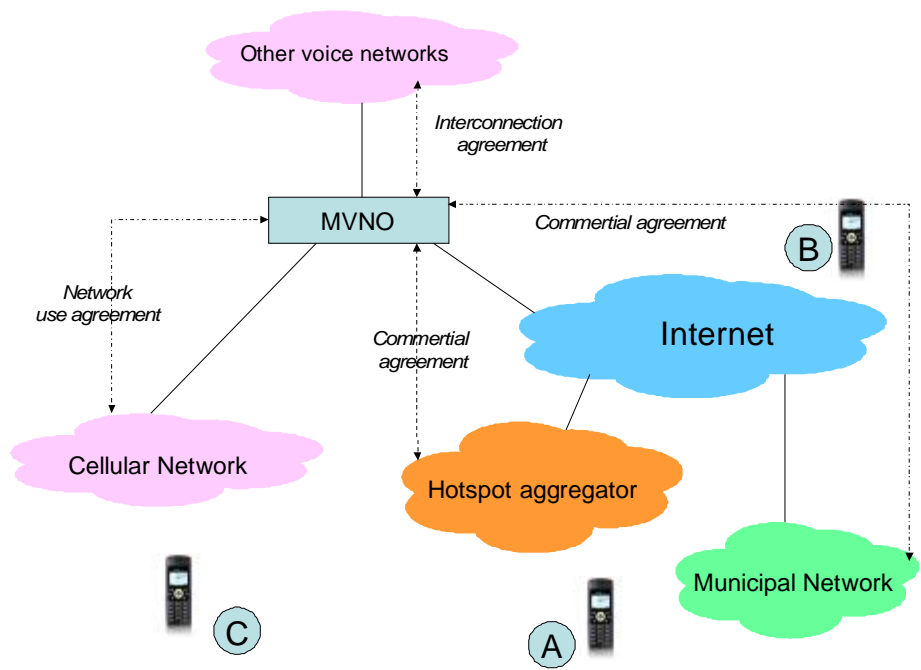

**Figure 7: MVNO combined cellular/WiFi VoIP Service**

In order to support this service, the MVNO would provide the user a VoIP E.164 non-geographic number to be used for both the outgoing and incoming calls and will receive and send calls via the cellular network only when the user is outside a WiFi coverage area. The MVNO is in this case able to manage relationship between the cellular network numbering and the VoIP one in order to hide to the client the complexity of having two E.164 numbers (VoIP and cellular one) associated.

As costs from using the cellular network are variable depending on the calls managed for the MVNO users, MVNO can benefit from users connected most of the time to WiFi networks, and share with the client the savings obtained, offering an integral coverage service (supported by the cellular network when the user is out of WiFi coverage) combined with a free call service for intra-Internet calls and low-cost service for outgoing calls to other voice networks.

As far as we know, no MVNO offers now this service. Regulatory barriers in many cases (non geographical numbering for VoIP services is not available in all countries), low maturity of MVNOs and dual terminals, difficulties on reaching interconnection agreements and the need to deploy signalling and transport infrastructure, has acted as slow down factors for configuring this business model.

As maturation will take place for adequate dual terminals availability, public WiFi networks coverage and technical and regulatory issues will be solved, it is likely that some MVNOs will probably start to test this type of business models, in collaboration with WiFi networks providers and pure VoIP service providers.

*Municipalities*

City or area wide WiFi networks are the most suitable deployments to compete directly with cellular networks, as they can provide continuous coverage in extent areas, solving at a local level the shortcoming on patchy coverage provided by the hotspots. The municipalities are the main actors that promote this dense deployment model, and they should considered the possible impact of providing these nomadic voice services on top of the municipal networks, probably in collaboration with other actors, such as services providers, local WISPs or fixed operators.

Considering that most of the mobile calls are done from the city or village where the user live, and supposing wide availability of dual WiFi/Cellular phones and a careful network design for supporting traffic at acceptable quality of service, many of the cellular calls could be handled by these dense WiFi municipal networks, deriving in profits for both the final users that would benefit from lower rates, and the municipal networks that would profit for more incomes, needed to make economically sustainable municipal networks deployment.

Also, the support of enhanced municipal services (both, internal and open to the citizens and visitors) based on combining voice and web-based information services, as well as nomadic voice for municipal officers and subcontracted personal, can reinforce the interest for these municipal networks.

To act as a real complement or competition to cellular networks, the interconnection with fixed and mobile telephony networks should be supported, and the user should be able to use the same number when under the municipal network coverage and when using the cellular service. In order to implement these interconnection and signaling capabilities, the municipalities will probably establish collaborations with VoIP service providers or MVNOs that will build this nomadic voice services on top of the municipal networks.

So, the strategy for municipal networks in to profit dense city coverage supplying in collaboration with specialized VoIP service providers a nomadic voice service inside the area where they supply WiFi service.

## 9. Evolution scenarios for voice services over WiFi

Based on the market actors strategies analyzed in the preceding section, some possible scenarios are shown and studied in order to identify to which extend could WiFi networks support voice over IP services. Each scenario is controlled by one or several of the market actors presented and has associated a time framework (short, medium or long term) when it is more likely to take place.

In the last part of the section, our view on how these scenarios will overlap or evolve is shown.

*Scenario 1: Cellular technologies success in retaining voice services*

As we have seen before, the technological decision for pure cellular operators is to bet for the evolution of 3G cellular networks and its evolution to increase nomadic bandwidth in order to support Internet and multimedia services in a convenient way to compete with other actors supporting their services in alternative infrastructures, as is the case for WiFi public networks.

The successful deployment of HSPA in many countries and the support and progressive maturation of LTE evolution to 4G points to a situation where 3,5G and 4G networks will be able to effectively compete with WiFi networks for most of the nomadic services ([Rysavy, 07]). The existing difference on bandwidth availability between cellular technologies and WiFi access networks will be in the medium and long term less relevant, terminal equipment will also in the long term equipped with both access technologies, and the practical differences for the user will be based on availability of coverage (ubiquitous on cellular networks) and rates (higher and minutes-based today for cellular networks).

In order to retain nomadic voice traffic, cellular operators should lower their mobile voice rates and will approach to flat-rate tariff schemes applying an evolution similar to the observed one on the voice fixed market, where the defensive strategy has been based on flat-rate call tariffs for national calls and bundling with in order to compete with VoIP service providers.

Although probably in the long term the price structure will evolve to flat-rate schemes, in the short term it is unlikely that the users will enjoy these not minute-based tarification, similar to the national fixed voice and broadband tariffs applied today. Cellular operators are not by now really threaded by alternative operators, being the only relevant competitors other cellular operators, and they have more variable costs depending on traffic than the ones supported by fixed operators.

While the existing minutes-based cellular tariffs will be still applied, there will exists incentives to enter the nomadic voice market for other alternative actors, based on lower rates for complementary service (to be used when available and using the integral coverage offered by cellular operators otherwise). Taking account that many of the cellular calls are made at home, at work or in places where WiFi coverage can be available, in can make sense for the user to profit VoIP WiFi service offers when available in order to alleviate nomadic voice bill.

In any case, cellular operators offer a convenient, simple to use, and integral coverage service, being difficult to compete with even applying lower tariffs, and the terminal equipment should evolve to incorporate as much simplicity and transparency as possible in order that WiFi public networks alone or in combination with other technologies will really act as a complement to cellular networks.

### *Scenario 2: Full Integrated fixed-cellular-WiFi voice service*

Integrated operators own both fixed high capillarity broadband networks as well as extense cellular 3/3,5G networks that can be combined in order to use a heterogeneous network to provide a pervasive transparent nomadic voice service.

The use of fixed infrastructure combined with WiFi access points for sending and receiving calls in places where the operator have this infrastructure available and shared with other services (as WiFi access points provided by operator in client premises) can help to optimize cellular infrastructure deployment, allowing for lower network costs than in the case of cellular operators.

As integrated operators can also provide transparent service and number integration when the user connects cellular, PSTN and Internet, they are in a good position to combine integral coverage when the user is not at home or near a hotspot (based on the cellular network), WiFi and PSTN connectivity when at home, and Internet-based connectivity when in any other place.

This capabilities, combined with the possibility of bundling a complete catalog of services (fixed and nomadic voice, Internet access, TV and multimedia services) in a transparent access technology base place integral operators in a advantageous position to compete with the rest of actors.

The case for these integral operators is similar to the pure cellular ones about opportunity windows for alternative actors. In the short and medium term prices for nomadic voice service will be still high leaving room for alternative VoIP WiFi business models.

### *Scenario 3: Competition based on integral WiFi municipal coverages*

This scenario is based on municipal networks succeed on finding sustainable models for deploying and exploiting wide area dense coverage WiFi networks. Combination of internal municipal services use and location-based publicity under collaborative models among local administrations and private initiative can enable in the medium term the provision of free VoIP WiFi services supported on municipal networks and low-cost calls to cellular and fixed networks.

In cities where these type of exploitation models will succeed, VoIP services supported on WiFi networks can absorb a relevant part of nomadic voice calls in places where reasonable coverage is provided. Nomadic services on top of municipal networks can also help to finance these local initiatives, making them more economically sustainable.

In the short term it is unlikely that this scenario will take place, as most of the municipal initiatives are not oriented to supply nomadic voice services but Internet browsing, e-mail and information services for citizens. In the medium term, the probability of success providing WiFi supported voice services depends on the ability to provide integral coverage in key areas of the city or village (a goal difficult and expensive to fulfill with the low-range ISM band WiFi technology).

Anyway, municipal networks providing a dense coverage in key areas of the city or village could act as a convenient complement to cellular networks for users combining WiFi in-home/in-work private network support with municipal network when in public domain locations as parks or commercial downtown and cellular service otherwise.

### *Scenario 4: Collaborative complement to cellular networks*

This scenario considers a success by fixed operators, hotspots aggregators and MVNOs to enter the nomadic voice market, acting as a convenient complement to cellular networks. In order to take place, they should establish alliances with VoIP service providers on top of the Internet to incorporate their large user base, and among them in order to supply as much as integral coverage and integrated service as possible.

As seen before, residential hotspots aggregators and fixed operators are starting to collaborate (FON alliances with BT and Time-Warner), but they are still oriented to data and information services, and VoIP services are considered as a nomadic service to be provided on these hotspots networks.

In order to offer a viable and convenient complement to cellular networks, the key actors will be in the future MVNOs, that are in a good position to coordinate this collaboration and manage heterogeneous network provided by residential hotspots aggregators in collaboration with fixed operators, municipalities, and commercial hotspots.

The scenario is complex, with many actors and not easy to be coordinated. Although in the short term it will probably not take place, a gradual establishment of alliances between actors can enable it in the medium term.

### *Scenario 5: Service Providers-driven complement to cellular networks*

As we have seen, WiFi networks are heterogeneous and controlled by a myriad of different actors (municipalities, small WISPs, aggregators, grassroots communities, traditional operators, etc.) that are difficult to be coordinated in order to supply a transparent one-stop-shopping service. Besides, pure service providers not owning networks and supplying service on top of the Internet are technology-agnostic, meaning that the service is provided on top of any access technology (being fixed, WiFi-based or even cellular networks) supporting the minimum requirements on QoS.

Considering these aspects, pure service providers not controlling networks, can offer their service without the need to collaborate with network actors and rely on the user for contracting access network services. In fact, this is the way that the most extended public VoIP service works: the user creates an account and operates on any network or combination of networks, not supplied by the service provider to access the service.

This scenario, then, is based on network and service layer separation: the service provider supplies the service on top of the Internet and the user contract or uses any network infrastructure accessible via WiFi. The service provider supplies a complement to cellular service and there is neither a total integration nor transparency on cellular and VoIP numbers.

As in this case the WiFi VoIP service is a natural extension of the fixed VoIP service, it is simple to be implemented for existing service providers that in fact do not care about the access technology.

As well as other cellular-complement strategies, is based on favorable tariffs (being usually free for communication among users adhered to the service) and can get a market share for users that most of the time have a WiFi network available for communication. In the long term, flat-rate tariffs for cellular or integral services under integral coverage will make this VoIP service less attractive.

In order to evolve the actual fixed VoIP service to integrate nomadic voice service and continue to extend the users adhered to the services, VoIP services providers should enjoy a favorable regulation for network neutrality. If network operators (being fixed, cellular or hotspots operators) identify VoIP services as a threat to their voice business and can legally restrict VoIP services provided by these actors, it will not possible to use third party networks to supply the service.

In any case, VoIP service providers are starting to establish collaborations with hotspots aggregators in order to facilitate nomadic VoIP service use in as much as locations as possible.

### *Is there a winner scenario for WiFi VoIP ?*

Provided technical issues are solved, such as the required QoS mechanisms to guarantee the bandwidth needed for a VoIP call, based on call admission control techniques, traffic differentiation / prioritization and other type of MAC (layer-2) enhancements to reduce the actual high overheads and improve the Hotspot capacity, there is no matter of doubt about the suitability of WLANs to carry VoIP already in the short / medium term.

Although in the short term most of the nomadic voice market will be still covered by cellular operators, we envisaged opportunities for alternative operators in order to get a market share in the nomadic voice services. These opportunities will be based in low-cost or free WiFi supported Voice over IP service acting as a non-transparent complement to cellular networks.

In this short term, the main actors that will try to get a share of the nomadic voice market supported by WiFi will be the existing VoIP service providers on top of Internet, as Skype. This VoIP service providers are establishing arrangements with WiFi networks providers as commercial and residential hotspots aggregators (Boingo-Skype, FON-Skype), and will probably also establish alliances with succeeding municipalities. The goal for VoIP providers is to extend the use of nomadic voice service in as much locations as possible and mobilize win-win strategies with WiFi networks operators, where both sides get a profit derived of the increased use of VoIP WiFi services. In the short term, the competition from VoIP WiFi supported services for the nomadic voice market will be based on free and low-cost service compared to the high prices for cellular service. This advantage will compensate the non-full transparent service provided by complementary actors to the cellular operators.

In the medium term, municipal wide area networks can play a relevant role to provide a dense coverage acting as a complement to cellular networks when the user is in the city. Also, more ubiquitous WiFi coverage can be obtained by collaboration among residential hotspots aggregators and fixed operators (as the case for BTFON), and both arranging agreements with municipal networks (as is the case for FON that are fixing agreements with municipalities in order to incorporate municipal coverage to their offer).

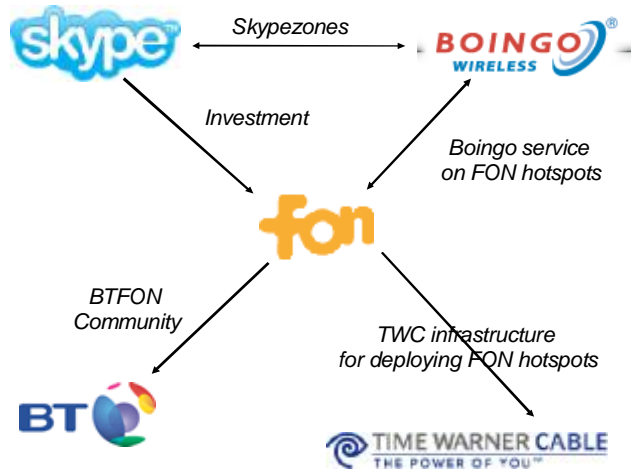

**Figure 8: Existing agreements among actors**

It is also expected that the use of VoIP over WiFi will also grow for at home use, steered by integral operators that will manage nomadic calls using dual phones that will used the fixed broadband infrastructure.

MVNOs combining WiFi dense coverage provided by municipalities, and residential coverage provided by aggregators as FON can play the role of nomadic voice integrators in the medium term, in collaboration with VoIP service providers enjoying large user base. As shown before, MVNOs deploying signalling and service management infrastructure could supply a transparent service similar to the one provided by the integral operators.

In the long run the evolution of cellular service be more and more based on flat-rate tariffs that will change the competition dynamics to fully integrated convenient nomadic ubiquitous services. Integral operators are in the best situation for optimizing network deployment and combining fixed, WiFi and cellular infrastructures in order to supply the service with the most adequate access technology for each call and location.

Succeeding actors in the short and medium term, as VoIP service providers on top of the Internet enlarging their user base, municipalities deploying city wide networks and hotspots aggregators reaching agreements with fixed operators can also maintain a part of the market share obtained in the short and medium term, establishing collaborations with MVNOs in order to provide a competitive nomadic transparent voice service.

So in fact, although today the market for nomadic voice is fully owned by cellular operators, there is a room for WiFi-based VoIP services, that will be based on a complementary low-cost not integrated with cellular service in the short term, driven by service providers in top of Internet. This complementary service will evolve to fully integrated, technology transparent service build on top of heterogenous cellular/fixed/WiFi networks, driven by integral operators in most of the cases or collaboration alliances among service providers, MVNOs, fixed operators, municipalities and aggregators succeeding in the competition with cellular networks in the short and medium term.

# 10. Conclusions

In spite of the irruption of new multimedia services, the voice service will continue to be the largest market in the future. Based on this assumption, in this paper we have evaluated if growing WiFi networks could get also a "piece of the cake" of the future nomadic voice market, competing with a numerous number of actors such as cellular and broadband cable telcos.

It's true that WLANs are not designed to carry voice. However, today, from the technical point of view, there are no reasons to not consider WLANs as a suitable access technology for wireless VoIP, especially now when the IEEE 802.11n standard promises transmission rates up to 600 Mbps, which means a capacity of hundreds of simultaneous calls in a single hotspot.

The position of WiFi service provides on this picture is not easy as it could look. Their main competitor will be the cellular telcos, which are evolving their infrastructure to high capacity enhanced 3G networks (now, based on HSPDA) and, although, they never will provide the huge bandwidth of WLANs, it would be more than enough for nomadic Internet access and VoIP. Additionally, there is the ubiquitous connectivity need, that cellular networks are able to guarantee, which is far away from the expected WiFi evolution in terms of integral coverage. Finally, the cheaper calls argument could be easily sorted from cellular networks if they evolve to flat-rate tariffs for voice.

However, the cellular network evolution (bandwidth provision) and the move to flat-rates still will require several years. For this period of time (short and medium term), there will be opportunities for WiFi networks based on: the role played by VoIP service providers such as Skype (based on price advantages), municipal networks supporting dense coverage in cities (tipically downtown areas), fixed operators extending telephony service to places where WiFi networks are available and MVNOs combining cellular and VoIP services in an integrated way.

Conversely, in the long term, it seems clear that: nomadic services will be provided by cellular networks or, at least, integrated heterogeneous infrastructures comprising cellular, fiber-based and WiFi technologies, but probably controlled by integral network operators. There is possible room for WiFi SPs growing enough (for example, from the integration of small WiFi initiatives in larger ones, in order to exploit scale economies) in the short and medium term and operating on top of the Internet under low-rates or free (publicity based or similar) service. In this case, access infrastructure is supplied by the traditional operators (layering), provided network neutrality is applied.

Therefore, an opportunity window in the short/medium term to act as a complement of cellular networks, supported on low/free rates compared to cellular. Several actors can simultaneously enter: VoIP SPs, municipal, fixed ops and MVNOs and alliances among them. In the long term, integral operators and cellular ones will continue to dominate the nomadic voice market, but if short/medium term alternative operators consolidate clients base, they can maintain their piece of the cake. So, yes, it can be an important application driven by integral operators in an heterogeneous network context in the long-run and can be also an important source of revenue in the medium term for WiFi networks providers and VoIP providers on top of Internet.

# 11. References

[Arjona, 07] ARJONA, A., TAKALA, S.: *"The Google Muni Wifi Network. Can it Compete with Cellular Voice?"*. The Third Advanced International Conference on Telecommunications, 2007. AICT 2007. Morne, Mauritius, 2007.

[Bellalta et al., 06] BELLALTA, B., MEO M**.:** *"Call Admission Control in WLANs"*. On book "Resource, Mobility and Security Management in Wireless Networks and Mobile Communications". Auerbach Publications, CRC Press, USA, September, 2006

[BTFon, 07] BTFON: *"BT and FON launch the world's largest Wi-Fi Community"*. [On line]. Available at <http://www.btfon.com/images/media/common/btfonLaunch041007.pdf>. [Last aAccessed: 30.10.2007].

[EC, 07] EUROPEAN COMMISSION: *"E-Communications household survey"*. Fieldwork November-December 2006. April, 2007. [On Line]. Available at
<www.ec.europa.eu/public_opinion/archives/ebs/ebs_274_sum_en.pdf>.[Last aAccessed: 21.08.2006].

[Dippon et al., 06] DIPPON, C., BANERJEE, A.: *"Mobile Virtual Network Operators: Blessing or curse?"*. Ed. by NERA. New York, 2006.

[ERG, 07] EUROPEAN REGULATORS GROUP: *"ERG common position on VoIP"*. December, 2007. [On line]. Available at
<www.erg.eu.int/doc/publications/consult_draft_cp_voip/erg_07_56_rev1_voip_draft_cp.pdf>. [Accessed: 08.01.2008].

[Felten, 06] FELTEN E.W.: *"Nuts and bolts of network neutrality".* The Journal of Policy, regulation and Strategy. July, 2006. [On line]. Available at <http://itpolicy.princeton.edu/pub/neutrality.pdf>. [Accessed: 19.07.2007].

[IEEE, 05] INSTITUTE OF ELECTRICAL AND ELECTRONICS ENGINEERS: *"Std 802.11e. Wireless LAN Medium Access Control (MAC) and Physical Layer (PHY) Specifications; Amendment: Medium Access Control(MAC) Quality of Service Enhancements"*. IEEE , 2005.

[Meinrath et al., 06] MEINRATH S.D., PICKARD V.W.: *" The New Network Neutrality: Criteria for Internet Freedom"*. October, 2006. 35th Telecommunications Policy Research Conference. George Mason University School of Law, Arlington, Virginia. [On Line]. Available at
<http://web.si.umich.edu/tprc/papers/2006/614/The%20New%20Network%20Neutrality%20v4.4.pdf>. [Accessed: 19.07.2007].

[Merrill Lynch, 04] MERRILL LYNCH: *"European wireless. Disruptive Technologies on the horizon?"*. March, 2004. [On Line]. Available at
<http://www.sims.berkeley.edu/~bigyale/Disruptive%20Technologies.pdf>. [Last accessed: 21.08.2006].

[Munasinghe, 08] MUNASINGHE, S., JAMALIPOUR, A.: *"Interworking of WLAN-UMTS Networks: An IMS-based Platform for Session Mobility"*. Accepted for publication in IEEE Communications, 2008.

[Muniwireless, 06] MUNIWIRELESS: *"State of the market report. 2006"*. October, 2006.

[Melody et al, 07] MELODY W.H., KELLY T.: *"The Challenges and Opportunities of VoIP"*. LIRNEasia Training Course on Strategies to Achieve Connectivity and Convergence. Singapur, 2007. [On line]. Availailable at <www.itu.int/osg/spu/presentations/2007/kelly-melody-challenges-opportunities-of-VoIP-1-march-07.pdf>. [Accessed: 01.09.2007] [Observer, 06] OBSERVER, THE: *"TeliaSonera launches the voice solution of the future"*. Press Note. August, 2006. [En LíneaOn line]. Available at <http://wpy.observer.se/wpyfs/00/00/00/00/00/08/29/F9/wkr0001.pdf >. [Last aAccessed: 29.08.2006].

[OCDE, 07] OCDE: *"OECD Communications Outlook 2007"*. ISBN 978-92-64-00681-2. [On line]. Available at < http://www.oecd.org/document/17/0,3343,en_2649_201185_38876369_1_1_1_1,00.html>. [Accessed: 22.07.2007].

[Odlyzko, 04] ODLYZKO, A.: *"Finding a voice: Learning from history"*. In "Connected Homes", Ed. Cisco. London, 2004 [On line]. Available at <http://www.cisco.com/web/about/ac79/docs/wp/ctd/CISCO_Connected_Homes.pdf>.[Accessed: 21.08.2006].

[Ofcom, 06] OFFICE OF COMMUNICATIONS: *"The International Communications Market 2006"*. Research document. November, 2006 [On line]. Available at <http://www.ofcom.org.uk/research/cm/icmr06/>. [Accessed: 06.12.2006]

[Ofcom, 07a] OFFICE OF COMMUNICATIONS: *"The International Communications Market 2006"*. Research document. August, 2007 [On Line]. Available at <http://www.ofcom.org.uk/research/cm/cmr07/>. [Last accessed: 01.09.2007]

[Ofcom, 07b] OFCOM: *"Research Report: Voice over Internet Protocol"*. July, 2007. [On line]. Available at < http://www.ofcom.org.uk/research/telecoms/reports/voip/>. [Accessed: 08.01.2008].

[Pujol, 05] PUJOL, F.: *"Wireless VoIP What threat to mobile operators?"*. Communications & Strategies Nº 60, 4th quarter 2005, p. 227. [On line]. Available at <www.idate.fr/fic/revue_telech/369/CS60_PUJOL.pdf >. [Accessed: 01.09.2007]

[Rysavy, 07] RYSAVY RESEARCH: *"EDGE, HSPA and LTE. The Mobile broadband advantage"*. September, 2007. [On line]. Available at < http://www.rysavy.com/Articles/2007_09_Rysavy_3GAmericas.pdf >. [Accessed: 06.11.2006].

[Samprakou, 05] SAMPRAKOU, I., BOURAS, C., KARAUBAILST, T.: *"Fast IP Handoff Support for VoIP and Multimedia Applications in 802.11 WLANs"*. World of Wireless Mobile and Multimedia Networks, Taormina - Giardini Naxos, 13-16 June 2005.

[Sfairopoulou et al., 08] SFAIROPOULOU, A., MACIÁN, C., BELLALTA B.: *"Adaptive codec selection for VoIP in multi-rate WLANs"*. On book "Wireless Multimedia: Quality of Service and Solutions". To be published by Idea Group Inc. June, 2008.

[Shamp, 04] SHAMP S.: *"Wifi clouds and zones: A Survey of Municipal Wireless Initiatives "*. New Media Institute and the Mobile Media Consortium. University of Georgia. [On Line]. Available at < http://www.nmi.uga.edu/research/WiFiCloudsZones--8-10-04.pdf>. [Accessed: 19.10.2004].

[Uma Technology, 06] UMA TECHNOLOGY: *"UMA Oververview"*. [On Line]. Available at < http://www.umatechnology.org/overview/index.htm>.[Accessed: 29.08.2006].

[Wi Revolution, 07] WI REVOLUTION: *"How does 802.11n get to 600Mbps?"* . September, 2007. [On Line]. Available at <www.wirevolution.com/2007/09/07/how-does-80211n-get-to-600mbps/> [Accessed: 26.04.2008]

[Wireless Philadelphia, 05] WIRELESS PHILADELPHIA EXECUTIVE COMMITEE: *"Wireless Philadelphia Business Plan"*. February, 2005, [On line]. Available at <www.phila.gov/wireless/pdfs/Wireless-Phila-Business-Plan-040305-1245pm.pdf>. [Accessed: 24.05.2005].